\documentclass[%
 aip,
 amsmath,amssymb,
reprint,%
]{revtex4-1}

\usepackage{amssymb}
\usepackage{graphicx}
\usepackage{dcolumn}
\usepackage{bm}
\usepackage[mathscr]{euscript}
\let\euscr\mathscr \let\mathscr\relax
\usepackage[scr]{rsfso}

\usepackage{stackengine}
\usepackage{xcolor}
\usepackage[utf8]{inputenc}
\usepackage[T1]{fontenc}
\usepackage{mathptmx}
\usepackage{etoolbox}
\usepackage{amsmath}
\usepackage{amsmath}
\let\oldAA\AA
\renewcommand{\AA}{\text{\normalfont\oldAA}}
\makeatletter
\def\@email#1#2{%
 \endgroup
 \patchcmd{\titleblock@produce}
  {\frontmatter@RRAPformat}
  {\frontmatter@RRAPformat{\produce@RRAP{*#1\href{mailto:#2}{#2}}}\frontmatter@RRAPformat}
  {}{}
}%
\makeatother

\begin{document}
\preprint{AIP/123-QED}

\title{GraphVAMPNet, using graph neural networks and variational approach to markov processes for dynamical modeling of biomolecules}

\author{Mahdi Ghorbani}
\email{ghorbani.mahdi73@gmail.com.}
\affiliation{ 
Laboratory of Computational Biology, National, Heart, Lung and Blood Institute, National Institutes of Health, Bethesda, Maryland 20824, USA.
}%
 \affiliation{Department of Chemical and Biomolecular Engineering, University of Maryland, College Park, MD 20742, USA}
 
\author{Samarjeet Prasad}%
\affiliation{ 
Laboratory of Computational Biology, National, Heart, Lung and Blood Institute, National Institutes of Health, Bethesda, Maryland 20824, USA.
}%

\author{Jeffery B. Klauda}
\affiliation{%
Department of Chemical and Biomolecular Engineering, University of Maryland, College Park, MD 20742, USA}%
\author{Bernard R. Brooks}%
\affiliation{ 
Laboratory of Computational Biology, National, Heart, Lung and Blood Institute, National Institutes of Health, Bethesda, Maryland 20824, USA.
}%

\date{\today}

\maketitle

Finding low dimensional representation of data from long-timescale trajectories of biomolecular processes such as protein-folding or ligand-receptor binding is of fundamental importance and kinetic models such as Markov modeling have proven useful in describing the kinetics of these systems. Recently, an unsupervised machine learning technique called VAMPNet was introduced to learn the low dimensional representation and linear dynamical model in an end-to-end manner. VAMPNet is based on variational approach to Markov processes (VAMP) and relies on neural networks to learn the coarse-grained dynamics. In this contribution, we combine VAMPNet and graph neural networks to generate an end-to-end framework to efficiently learn high-level dynamics and metastable states from the long-timescale molecular dynamics trajectories.  This method bears the advantages of graph representation learning and uses graph message passing operations to generate an embedding for each datapoint which is used in the VAMPNet to generate a coarse-grained representation. This type of molecular representation results in a higher resolution and more interpretable Markov model than the standard VAMPNet enabling a more detailed kinetic study of the biomolecular processes. Our GraphVAMPNet approach is also enhanced with an attention mechanism to find the important residues for classification into different metastable states.

\section{Introduction}

Recent advances in computer hardware and software has recently enabled the generation of extensive and high throughput molecular dynamics (MD) trajectories.\cite{lindorff2011fast,shaw2021anton} These facilitates the thermodynamic and kinetic study of biomolecular processes such as protein folding, protein-ligand binding and conformational dynamics to name a few. These simulations often produce large amount of high dimensional data which require special rigorous techniques for analyzing the thermodynamics and kinetics of the molecular processes.  

In recent years, Markov state modeling approach \cite{swope2004describing, husic2018markov, mcgibbon2015variational, prinz2011markov} has been greatly developed and utilized for understanding long-timescale behavior of dynamical systems and state-of-the art software packages such as Pyemma\cite{scherer2015pyemma} and MSMBuilder \cite{harrigan2017msmbuilder} were introduced. Markov state models provide a master equation that describes the dynamic evolution of the system using a simple transition matrix.\cite{scherer2015pyemma} Markovianity in these system means the kinetics are modeled by memoryless jumps between  states in the state space. Combined with the advances in the MD simulations, the framework for MSM construction has been greatly advanced to a robust set of methods to analyze a dynamical system. In an MSM the molecular conformation space is discretized into coarse-grain states, where the inter-conversions between microstates within a macrostate are fast compared to transitions between different macrostates.\cite{sidky2019high} Markov state models have previously been used to investigate kinetics and thermodynamic properties of biophysical systems such as protein folding \cite{noe2009constructing, bowman2009progress, voelz2010molecular, beauchamp2012simple}, protein-ligand \cite{held2011mechanisms, plattner2015protein} binding and protein conformational changes.\cite{kohlhoff2014cloud, banerjee2015conformational, sadiq2012kinetic}

There are several steps in the pipeline of Markov model construction: The first step involves featurization where relevant MD coordinates such as distances, contact maps or torsion angles are chosen.\cite{mcgibbon2015mdtraj} This is followed by a dimension reduction step that maintains the slow collective variables using methods such as time independent component analysis (TICA)\cite{perez2013identification, schwantes2013improvements} or dynamic mode decomposition (DMD) \cite{mezic2005spectral, schmid2010dynamic, tu2013dynamic} or other variants of these techniques. The resulting low-dimensional space is then discretized into discrete states.\cite{chodera2007automatic, wu2015gaussian, weber2017set} This is usually done using Kmeans clustering.\cite{scherer2015pyemma} A transition matrix is then built on the discretized trajectories which describes the time evolution of processes using a lagtime. \cite{prinz2011markov, bowman2009progress} This transition matrix can be further processed through its Eigendecomposition to find the equilibrium and kinetic properties of the system.\cite{prinz2011markov} Finally, fuzzy clustering methods such as PCCA are often used to produce a more interpretable coarse-grained model.\cite{roblitz2013fuzzy} 

As noted above, there are multiple steps where hyperparameters must be carefully chosen to construct the Markov model. The quality of the constructed MSM is highly dependent on these steps and this has brought many research opportunities into optimizing the pipeline of MSM using various techniques.\cite{prinz2011markov, noe2013variational, schwantes2013improvements, husic2017ward, mcgibbon2015variational, husic2016optimized, scherer2019variational} Moreover, complex dynamical systems require the optimal choice of model parameters which requires physical and chemical intuition about the model. Therefore, suboptimal choices of model parameters can lead to poor results in learning the dynamics from trajectory. Recently a variational approach for conformational dynamics (VAC) has been proposed which help in selection of optimal Markov models by defining a score that measures the optimality of the given Markov model for governing the true kinetics.\cite{wu2017variational,noe2013variational,scherer2019variational,nuske2014variational,scherer2019variational,mcgibbon2015variational} VAC states that given a set of n orthogonal functions of state space, their time autocorrelations at lag-time $\tau$ are the lower bounds to the true eigenvalues of the Markov operator. \cite{nuske2014variational} This is equivalent to underestimating the relaxation time scales and overestimating the relaxation rates.\cite{wu2017variational, noe2013variational, nuske2014variational} Before VAC, the tools to diagnose the performance of MSMs were mainly visual such as implied timescale plot (ITS)\cite{swope2004describing} and Chapman-Kolomogorov test\cite{husic2018markov} (CK-test). Variational approach enabled the objective comparison of different model choices for the same lag time. \cite{noe2013variational} The VAC has recently been generalized to variational approach for Markov processes (VAMP).\cite{wu2017variational,wu2020variational} VAMP was proposed for the general case of irreversible and non-stationary time series data and is based on a singular value decomposition of the Koopman operator.\cite{wu2020variational}   
\raggedbottom

Using this variational principle, Mardt and coworkers introduced VAMPNets to replace the whole pipeline of Markov model construction with a deep learning framework.\cite{mardt2018vampnets} A VAMPNet maps the configuration $x$ to a low-dimensional fuzzy state space where each timepoint has a membership probability to each of the metastable states. VAMPNets are then optimized by a variational score (VAMP-2). This framework was further developed by directly learning the eigenfuctions of the spectral decomposition of the transfer operator that propagates equilibrium probability distribution in time with a state free reversible VAMPNet (SRV).\cite{chen2019nonlinear} Physical constraints of reversibility and stochasticity were further added to the model to have a valid transition matrix enabling the computations of transition rates and other equilibrium properties from out-of-equilibrium simulations.\cite{mardt2020deep, mardt2021progress} However, a proper representation of the molecules is not discussed in these models and traditional distance matrices or contact maps are often used.

Conformational heterogeneity of proteins during folding can complicate the selection of features for building a dynamical model. This is even more true in the case of disordered proteins.\cite{lohr2021kinetic} Gaining interpretable few state kinetic model of protein folding using MD trajectories is still highly desirable and can be achieved by MSM approaches using carefully chosen features. For this, one would choose a set of features for the system such as distances, dihedral angels or the root-mean-square deviation to some reference structure. Using more complicated feature functions such as convolutional layers on the distance matrices was proposed to enhance the kinetic resolution of the model.\cite{lohr2021kinetic} Graph neural network have previously been used in molecular feature representation as a promising tool in a variety of applications to predict the properties of the system of interest or energies and forces.\cite{schutt2017schnet,husic2020coarse} Battaglia first introduced graph neural networks with convolutional operations and graph message passing.\cite{battaglia2018relational,kipf2016semi} Currently there are various types of graph neural networks that differ in how the message passing operations are done between nodes and edges and how the output of the network is generated. Traditionally, distance maps or contact maps were used to represent the structure of the molecules. A more natural way of representing proteins is by using graphs where nodes represents atoms and the edges representing the bonds (real or unreal) connecting them. This representation is rotationally invariant by construction. Recent advances of protein structure prediction has greatly exploited the advances in geometric deep learning \cite{bronstein2017geometric} and graph neural networks \cite{kipf2016semi, bruna2013spectral}. Combining  the VAMPNet framework with graph neural network improves the kinetic resolution of the resulting low dimensional model where a smaller lagtime can be chosen to build the transition matrix. In the original VAMPNet the dynamics is directly coarse-grained into a few coarse-grained states without learning a low-dimensional latent space representation. However, using graph neural networks, we show that the learned embeddings for the graphs can represent useful information about the dynamic system. Furthermore, using a graph attention network \cite{velivckovic2017graph} gives us useful insights about the importance of different nodes and edges for different metastable states.

\begin{figure*}
    \centering
    \includegraphics[scale=0.49]{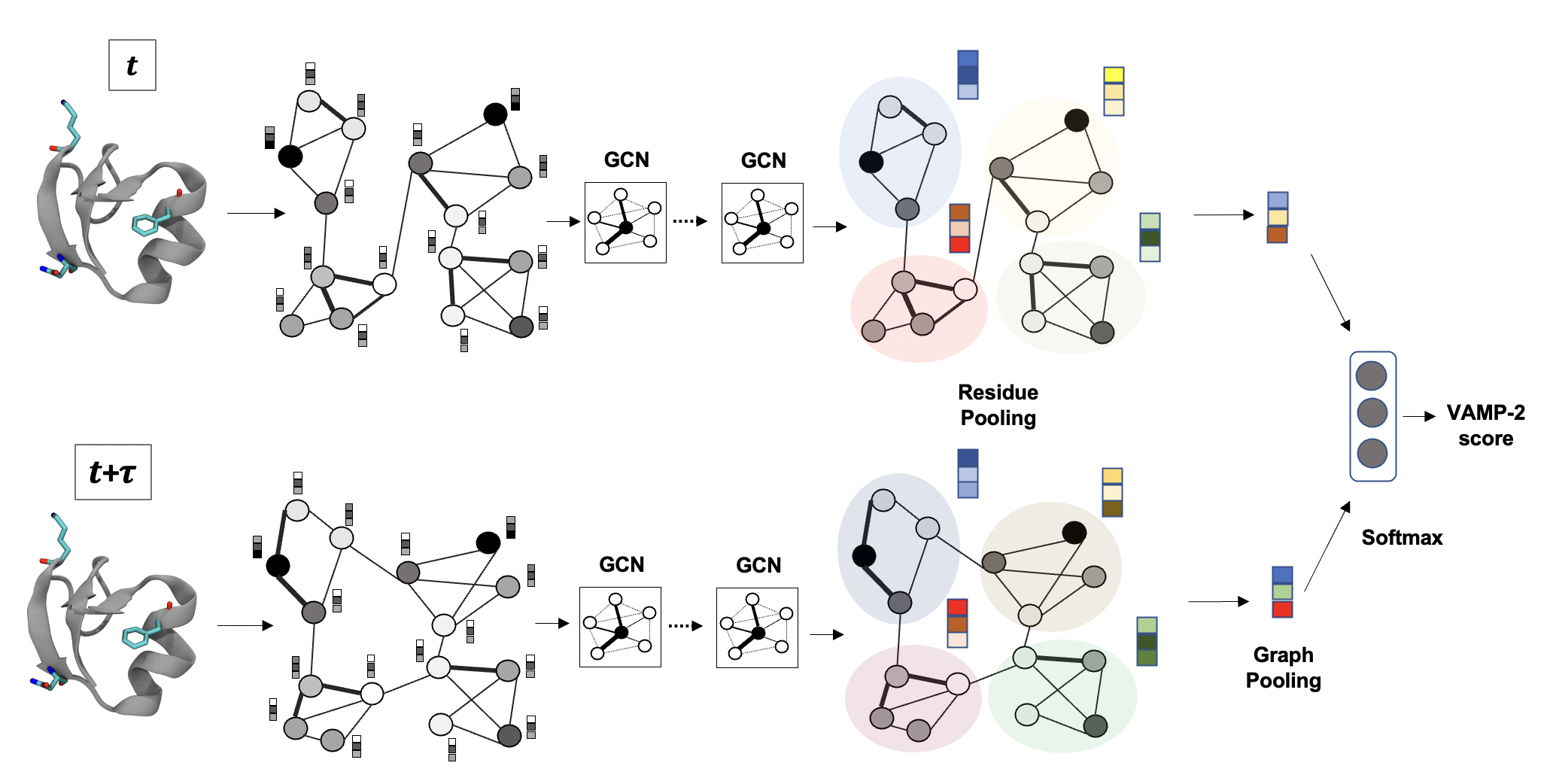}
    \caption{Overview of the architecture of GraphVAMPNet method. Given a molecular structure at time $t$ and a lagtime later $t+\tau$ , molecular graphs are built using the nearest neighbor of the chosen atoms. Several graph convolution operations are performed resulting in representation for each node. A hierarchical pooling is done to find a latent representation of the full graph which is concatenated between time $t$  and $t+\tau$. The full network is then optimized by maximizing a VAMP-2 score.}
    \label{fig:my_label}
\end{figure*}

\section{Methods}

A Markov model estimates the dynamics by a transition density which is the probability density of transition to state $y$ at time $t+\tau$ given that the system was at state $x$ at time $t$:

\begin{equation}
    p_\tau (x,y) = P(x_{t+\tau} = y | x_t = x)
\end{equation}

Where $x$ and $y$ are two different states of the system and $\tau$ is the lagtime of the model from which the transition probability density $P$ is built. Using this definition of transition density the time evolution of the ensemble of states in the system can be written as:

\begin{equation}
  p_{t+\tau}(y) = (\euscr{P}_\tau p_t ) (y) = \int {p_\tau (x,y) p(x) dx} 
\end{equation}
In this equation $P_\tau$ acts as a propagator which propagates the dynamics of the system in time. However, this definition of propagator assumes a reversible and stationary dynamical systems.\cite{prinz2011markov} For the general case of non-reversible and non-stationary dynamics, Koopman operator is used.\cite{wu2017variational} Koopman theory enables feature transformations into a feature space where the dynamics evolve on average linearly. Koopman operator acts like a transition matrix for non-linear dynamics and describes the conditional future expectation values for a fixed lag time $\tau$. In the Koopman theory the Markov dynamics at a lagtime $\tau$ is approximated by a linear model of the following form:

\begin{equation}
    \mathbb{E}[X_1(x_{t+\tau})] \approx K^T \mathbb{E}[X_0(x_t)]
\end{equation}

In equation above $X_0(x)=(X_{00} (x), ... X_{0m} (x)^T$ and $X_{1}(x)=(X_{11} (x), ..., X_{1m}(x)^T$ are feature transformations to a space where dynamics evolve on average linearly. This approximations is exact in the infinite-dimensional feature transformations however it was shown that given a large enough lagtime $\tau$ the low dimensional feature transformations can become optimal.\cite{wu2017variational}  Equation 3 can be interpreted as a finite rank approximation of the so-called Koopman operator.\cite{mezic2013analysis} The optimal Koopman matrix to minimize the regression error from equation 3 is:

\begin{equation}
    K = C_{00}^{-1} C_{01}
\end{equation}

Where the mean-free covariance matrices for data transformation are defined as:

\begin{equation}
\begin{cases}
C_{00} = \mathbb{E} [X_0(x_t)X_0(x_t)^T] \\
  C_{0t} =\mathbb{E} [X_0(x_t)X_1(x_{t+\tau})^T]  \\
C_{tt} = \mathbb{E} [X_1(x_{t+\tau})X_1(x_{t+\tau})^T]
\end{cases} 
\end{equation}

However, the regression error has no information about the choice of feature transformations $X_0$ and $X_1$ and can lead to trivial solutions for these feature transformations.\cite{mardt2018vampnets} On the other hand, VAMP provides useful scoring functions that can be used to find optimal feature transformations. VAMP is based on singular value decomposition of Koopman operator and is used to optimize the feature functions and does not have the limitations of time reversibility and stationary dynamics. VAMP states that given a set of orthogonal candidate functions, their time-autocorrelations are the lower bounds to the true Koopman eigenvalues. This provides a variational score such as the sum of estimated eigenvalues that can be maximized to find the optimal kinetic model. Wu and Noe showed that the optimal choice of $X_0$ and $X_1$ in equation 3 are obtained using the singular value decomposition of the Koopman matrix and setting $X_0$ and $X_1$ to its top left and right singular functions respectively.\cite{wu2017variational, wu2020variational}

The VAMP-2 score is then defined as :
\begin{equation}
    VAMP-2 = \sum \limits_i \sigma_i ^2 = || C_{00} ^{-1/2} C_{0t} C_{tt} ^{-1/2} ||_F ^ 2 + 1
\end{equation}

The left and right singular functions of the Koopman operator are always equal to the constant function 1. Therefore, 1 is added to basis functions. Maximum VAMP-2 score is achieved when the top m left and right Koopman singular functions are in the span  ($X_{01}$, ...$X_{0m}$) and ($X_{11}$, ..., $X_{1m}$) respectively. VAMP-2 also maximizes the kinetic variance captured by the model.

Feature transformations $X_0$ and $X_1$ can be learned with neural network in the so called VAMPNet where there are 2 parallel lobes each receiving MD configurations $x_t$ and $x_{t+\tau}$. As done in the original VAMPNet, we assume the two lobes have similar parameters and use a unique basis set $X=X_0=X_1$. The training is done by maximizing the VAMP-2 score to learn the low-dim state space produced by a softmax function. 
Since K is a Markovian model it is expected to fulfill the Chapman-Kolmogrov (CK) equation
\begin{equation}
    K(n\tau) = K^n (\tau)
\end{equation}

for any value n>=1 where $K(\tau)$ and $K(n\tau)$ indicate models estimated at a lag time of $\tau$ and $n\tau$ respectively. 

The implied timescales of the process are computed as follows:

\begin{equation}
    t_i (\tau) = - \frac{\tau}{ln|\lambda_i (\tau)| }
\end{equation}
    
Where $\lambda_i (\tau)$ is the eigenvalue of the Koopman matrix built at a lagtime $\tau$. The smallest lagtime $\tau$ is chosen where the implied timescales $t_i(\tau)$ are approximately constant in $\tau$. After having chosen the lagtime $\tau$ we test whether the CK test holds within statistical uncertainty.

\subsection{Protein Graph representation}
Each structure is represented in terms of an attributed graph $G=(V,E)$ where $V$ are node features $V={v_1, ..., v_N}$ and $E$ are the edge features $E={e_{ij}}$ that captures the relations between nodes. We have tested different Graph Neural networks (GNNs) for their ability to learn higher resolution kinetic models from MD trajectories of protein folding simulations.\cite{xie2019graph} In all these different GNNs the node embeddings $v_i$ are initialized randomly and the edge embeddings $e_{ij}$ are the Gaussian expanded distances between the adjacent nodes using the following formula:
\begin{equation}
    e_{ij}^t = exp(-(d_{ij} - \mu_t )^2/\sigma ^2)
\end{equation}
Where $d_{ij}$ is the distance between atoms $i$ and $j$, $\mu_t = d_{min} + t\times(d_{max} - d_{min})/K$ $\AA$, $t=0,1,...,K$,  $\sigma=(d_{max}-d_{min})/K$ $\AA$. $d_{max}$ and $d_{min}$ are the maximum and minimum distances respectively for constructing the gaussian expanded edge features.  Unless noted otherwise, all the graphs are built using M nearest neighbors for the $C_{\alpha}$ atoms of the protein with edges built from the gaussian expanded distances between $C_{\alpha}$ atoms. 

\subsubsection{Graph Convolution layer}
In this type of graph neural network, protein graph is represented as $G=(V,E)$ where $V$ contains features of the nodes and $E$ contains the edge attributes of the graph. A separate graph is constructed for configurations at each timestep of the simulation. The initial node feature representations are randomly initialized. However, a one-hot vector representation based on the atom type or the amino acid type can also be used. During training the node embeddings $v_i ^k$ for node $i$ at layer $k$ are updated using the following equations:\cite{sanyal2020proteingcn, xie2018crystal}

\begin{equation}
    v_{i}^{k+1} = v_{i}^{k} + \sum\limits_{j\in N_i} w_{i,j}^k \odot g(z_{i,j}^{k} W_{c}^k + b_{c}^{k})
\end{equation}
\begin{equation}
    w_{i,j}^{k} = \sigma (z_{i,j}^{k} W_{g}^{k} + b_{g}^{k} ) 
\end{equation}
\begin{equation}
    z_{i,j}^{k} = v_{i}^k \oplus v_{j}^k \oplus e_{i,j}
\end{equation}

Where $\odot$ denotes element-wise multiplication and $\oplus$ denotes concatenation, $\sigma$ is the sigmoid function as the non-linearity and $g()$ is the edge-gating mechanism introduced by Marcheggiani\cite{marcheggiani2017encoding} to incorporate different interaction strength among neighbors into the model. $W_{g}^k$, $W_{c}^k$, $b_{g}^k$ and $b_{c}^k$ are gate weight matrix, convolution weight matrix, gate bias and convolution bias respectively for k'th layer of the graph convolution layer. To capture the embedding of the whole graph, we use graph pooling where graph embeddings are generated using the learnt node embeddings.\cite{hamilton2017inductive, ranjan2020asap} The embedding for the whole graph is done through a pooling function where we average over the embedding of all the nodes.

\begin{equation}
    v_G = \frac{1}{N} \sum_{i=1}^{N}{v_i}
\end{equation}
Other types of pooling such as hierarchical pooling can also be applied to make a more complicated model.\cite{ying2018hierarchical}

\subsubsection{SchNet}

Another type of GNN is SchNet which was introduced by Schutt and others to use continuous-filter convolutions for predicting forces and energies of small molecules according to quantum mechanical models.\cite{schutt2017schnet, battaglia2018relational}. This was modified by Husic et al.\cite{husic2020coarse} to learn a coarse-grained representation of molecules using graph Neural network. SchNet is employed here to learn feature representation of nodes for learning dynamics of protein folding. This is a subunit of our model for learning feature representations of molecules on the graph level. The initial node features or embeddings are initialized randomly but a one-hot encoding based on node type (amino acid) can also be used. These embeddings are later learned and updated during training of the network by a few rounds of message passing through nodes and edges of the graph. Node embeddings are updated in multiple interaction blocks as implemented in the original SchNet.\cite{schutt2017schnet} Each interaction layer contains continuous convolution between nodes. The edge attributes are obtained by using a radial basis function e.g. Gaussians centered at different distances. 

\begin{equation}
    e_{ij} = exp(-\gamma (d_{ij} - \mu )^2)
\end{equation}

These edge attributes are then fed into a filter-generating network $w$ that maps $e_{ij}$ to a $d_h$-dimensional filter. This filter is then applied to node embeddings as (continuous filter convolution):

\begin{equation}
    z_{i}^k = \sum\limits_{j\in N(i)} \alpha_{ij} ^ k w^k e_{ij} \odot b^k (h_{i}^k) 
\end{equation}

\begin{equation}
    \alpha_{ij}^k = \frac{exp(z_{ij}^k W_a ^k)}{\sum_j exp(z_{ij}^k W_a ^t)}
\end{equation}

\begin{equation}
   z_{ij}^k = {\bigoplus} w^k e_ij \odot b^k (h_i ^k)
\end{equation}
$w$ is a dense neural network and b is an atom-wise linear layer as noted in the original paper.\cite{schutt2017schnet} Note that the sum is over every atom $j$ in the neighborhood of atom $i$. Multiple interaction blocks allow all atoms to interact with each other in the network and therefore allow the model to express complex multi-body interactions. We enhanced the standard SchNet architecture by adding an attention layer that learns the importance of the edge embeddings for updating the embedding of incoming node in the next layer. The attention weight $\alpha_{ij}$ is learned using a softmax function between embedding of the node and its neighbors. $\oplus$ denotes the concatenation of node features of neighboring nodes $j \in N(i)$ where $i$ is the query node. The node embeddings are updated in each interaction block, which can contain a residual block to avoid gradient annihilation as done in deep neural networks.\cite{he2016deep} The residual connection is followed by a nonlinear activation function of the output of continuous filter convolution $z_{i}^k$ as:
\begin{equation}
    h_{i}^{k+1} = h_{i}^{k} + g^{k}(z_{i}^k)
\end{equation}
The trainable function g involves linear layers and a nonlinearity. We used a hyperbolic tangent as the activation as proposed by Husic et al.\cite{husic2020coarse} The output of final SchNet interaction block is fed into an atom-wise dense network. The learned embedding of nodes after several SchNet layers is then fed into a pooling layer as described previously to produce a graph embedding for each timestep.

\subsection{Model selection and Hyperparameters}

In GraphVAMPNet, instead of traditional features such as dihedral angles, distance matrices and contact maps, we use a general graph neural network, a more natural representation of molecules and proteins. We have implemented two different graph neural networks (GraphConvLayer and SchNet). The GraphVAMPNet built from each of these GNN layers have several model hyperparameters including the dimension of feature space (number of output states) and the lag time $\tau$. To resolve $k-1$ relaxation timescales, we need at least $k$ output neurons in the last layer of the network since the softmax function removes one degree of freedom. The models are trained by maximizing the VAMP-2 score on the training set and hyperparameters are optimized using a a cross-validated VAMP-2 score for the validation set using a ratio of 0.7 for training and 0.3 for the validation set. To have a fair comparison between different feature representations we trained model with similar number of layers (4) and similar number of neurons per layer (16). In general, increasing the dimension of feature space makes the dynamical model more accurate, but it may result in overfitting when the dimension is very large. A higher dimensional feature space is also harder to interpret as the model seeks a low dimensional representation. Therefore, in this study, we experiment on systems with 5-state output models unless stated otherwise. There are multiple hyperparameters in the model that must be selected. These include the architecture of GCN, the number of clusters, number of neighbors for making the graph, time step of analyzing the simulation. Here we have used 5 clusters for Trp-Cage and NTL9 and 4 clusters for Villin. In the case of Villin using a 5-state led to finding only 4 states after using a cutoff value of 0.95 on the cluster probabilities which is the reason we have used a value of 4 for this protein. The hyperparameters chosen for each protein are shown in table 1.
\begin{table*}
\caption{Hyperparameters for each system in this study}
\begin{tabular}{ccccccccc}
\hline
\textbf{system}  & \textbf{\begin{tabular}[c]{@{}c@{}}number\\ of graph\\ layers\end{tabular}} & \textbf{\begin{tabular}[c]{@{}c@{}}number \\ of \\ neurons\end{tabular}} & \textbf{\begin{tabular}[c]{@{}c@{}}number \\ of \\ clusters\end{tabular}} & \textbf{Batch-size} & \textbf{\begin{tabular}[c]{@{}c@{}}learning\\ rate\end{tabular}} & \textbf{\begin{tabular}[c]{@{}c@{}}number\\ of\\ atoms\end{tabular}} & \textbf{\begin{tabular}[c]{@{}c@{}}number\\ of\\ neighbors\end{tabular}} & \textbf{\begin{tabular}[c]{@{}c@{}}number\\ of\\ Gaussians\end{tabular}} \\ \hline
\textbf{TrpCage} & 4                                                                           & 16                                                                       & 5                                                                         & 1000                & 0.0005                                                           & 20                                                                   & 7                                                                        & 16                                                                       \\ \hline
\textbf{Villin}  & 4                                                                           & 16                                                                       & 4                                                                         & 1000                & 0.0005                                                           & 35                                                                   & 10                                                                       & 16                                                                       \\ \hline
\textbf{NTL9}    & 4                                                                           & 16                                                                       & 5                                                                         & 1000                & 0.0005                                                           & 39                                                                   & 10                                                                       & 16                                                                       \\ \hline
\end{tabular}
\end{table*}

\begin{figure*}
    \centering
    \includegraphics[scale=0.55]{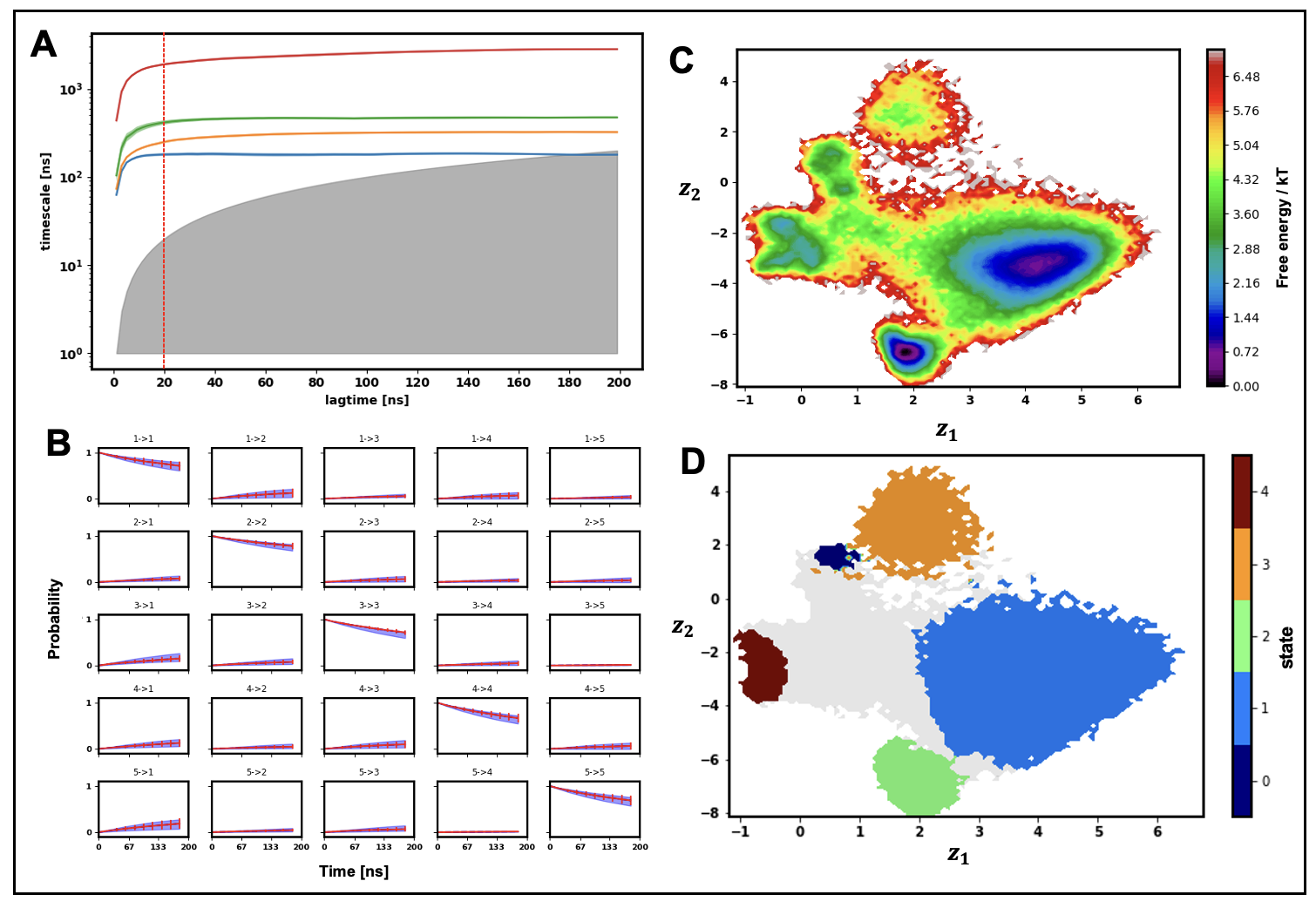}
    \caption{ TrpCage system A) Implied timescale (ITS) plot for SchNet as feature transformation in VAMPNet (errors are taken from 95 \% confidence interval from 10 different trainings) B) CK-test for SchNet using lagtime of 20 ns C)  Free energy landscape (FEL) in a 2d graph embedding F) state assignment of the 2d graph embedding using 0.95 cutoff. }
    \label{fig:my_label}
\end{figure*}

\section{Results}
We tested the performance of our GraphVampNet method on 3 different protein folding systems including Trp-Cage (pdb: 2JOF)\cite{barua2008trp}, Villin (pdb: 2F4K)\cite{kubelka2006sub} and NTL9 (pdb: 2hba)\cite{cho2014energetically}.  The Graph Neural network was implemented using PyTorch and the deeptime \cite{hoffmann2021deeptime} package was used for VAMPNet. Pyemma\cite{scherer2015pyemma} was used for free energy landscape plots. Adam was used as the optimizer in all models. GNN provides a framework to learn the feature transformations in VAMPNets that is invariant to permutation, rotation and reflection.  Moreover, the graph embeddings can be learned with the GraphVAMPNet framework which correspond to different dynamic states during the simulation. In order to visualize the graph embeddings in 2-D we have also transformed the graph embeddings in the last layer of GraphVAMPNet into 2-D and trained the model by maximizing the VAMP-2 score. The free energy landscape on the graph embeddings shows highly separated metastable states separated by high energy transition regions. The low energy metastable states correspond to the regions of high fidelity for metastable assignment probabilities with higher than 0.95. It is important to note that this is only done for visualization purposes and higher dimensional (16) embeddings are used for finding the metastable states in these complex protein folding systems. Furthermore, the present results do not depend on enforcing reversibility to the learned transition matrix. However, this can be done by Koopman re-weighting\cite{wu2017variational} or learning the re-weighting vectors during training in the VAMPNet framework.\cite{mardt2020deep}

Although we have tried different Graph Convolution networks in the GraphVAMPNet approach, our results showed that SchNet has the best performance with the highest VAMP-2 score among all. Therefore, we present the results of SchNet in the main paper. The average VAMP-2 scores are calculated from the validation set 10 different training for each system and compared (table S1). The VAMP-2 score for SchNet in each system for training and validation set are plotted against the training epoch and shows a converging behavior after 100 epochs (figure S1).

\subsection{Trpcage}

We test our GraphVAMPNet model an ultra-long 208 $\mu s$ explicit solvent simulation of K8A mutation of the 20-residue Trp-Cage TC10b at 290 K provided by DE Shaw group.\cite{lindorff2011fast} The folded state of Trp-Cage contains an $\alpha$-helix (residues 2-8), a $3_{10}$-helix and a polyproline II helix.\cite{meuzelaar2013folding} The tryptophan residue (Trp6) is caged at the center of the protein.

A VAMPNet was built for different types of feature learning in neural networks. The average VAMP-2 score of the validation set for 10 training runs were compared between different types of feature learning Neural Networks (Standard VAMPNet, and two graph layers) in table S1. SchNet showed the highest average VAMP-2 score among different types of features leanings. The average VAMP-2 score for training and validation set of TrpCage for 10 different training examples is shown in figure S1A. The VAMP-2 score shows a converging behavior after 100 epochs of training. Since SchNet showed the highest VAMP-2 score we use this type of Graph Neural network for the rest of our analysis.  The implied timescales learned using SchNet is shown in figure 2A. Implied timescale plots for GraphConvLayer and standard VAMPNet are shown in figures S2A,B. Standard VAMPNet using distances shows higher variance in the implied timescales than the VAMPNet built using SchNet. A faster convergence of implied timescales for SchNet is also observed compared to standard VAMPNet. All 4 timescales in SchNet converge after 20 ns which is the lagtime we choose to build the kinetic Koopman matrix.  Moreover, a closer look at the implied timescales shows that the timescales learnt from standard VAMPNet layer are also smaller (e.g. the fourth timescale) than the implied timescales in SchNet. According to the variational approach for Markov processes, a model with longer implied timescale corresponds to less modeling error of the true dynamics of the system.\cite{nuske2014variational} To validate the resulting GraphVAMPNet, we conduct a CK-test which compares the transition probability between pairs of states $i \rightarrow j$ at time $k\tau$ predicted by a model at lagtime $\tau$. CK-test shows (figure 2B) excellent prediction of transition probabilities even at large timescales $k\tau=200$ ns at a lagtime of 20 ns. 
\begin{figure*}
    \centering
    \includegraphics[scale=0.6]{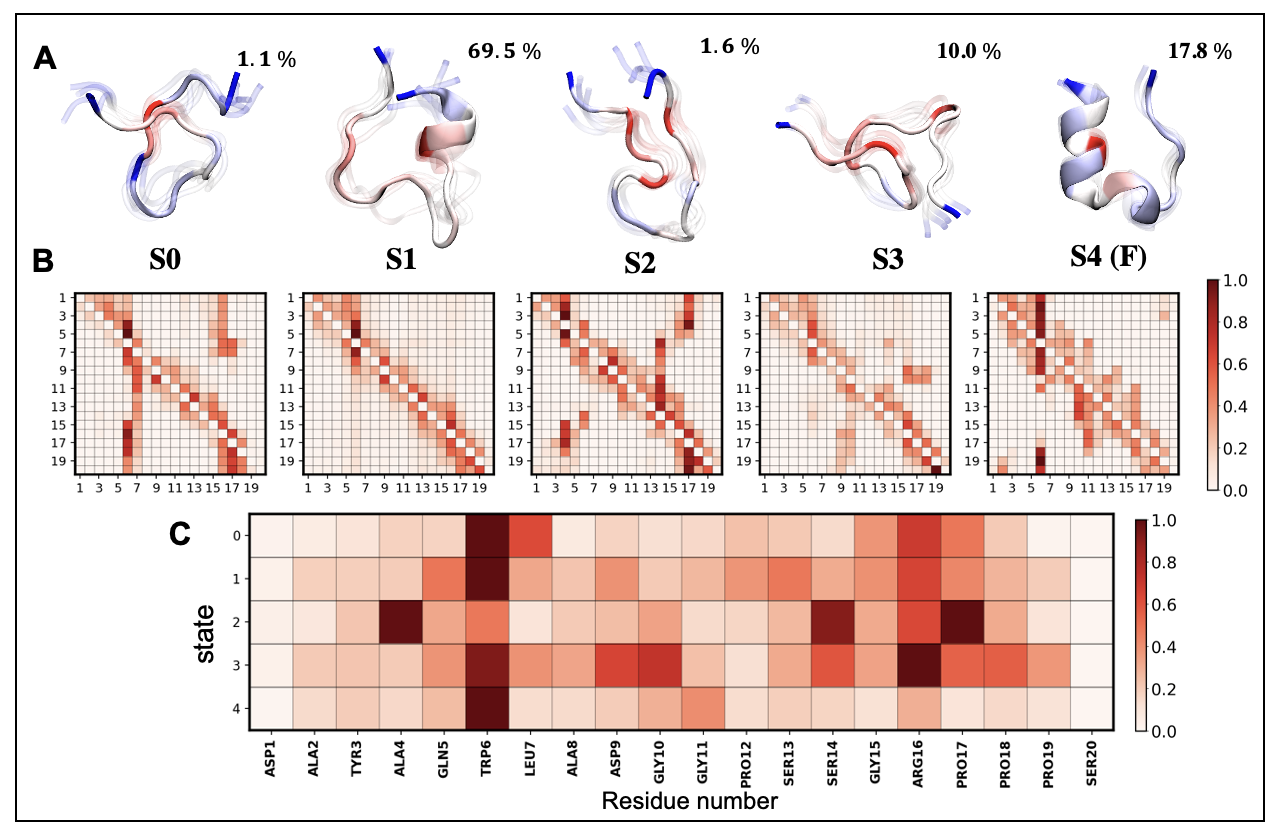}
    \caption{A) Representative structure of each metastable state in TrpCage with their probabilities B) average attention score between $C_\alpha$ atoms for each cluster C) averaged attention score for each residue of TrpCage in each cluster which is the scaled sum of rows.}
    \label{fig:my_label}
\end{figure*}
We next analyzed the resulting coarse-grained states built from VAMPNet using SchNet as feature transformation. The folded state (S4) possesses 18$\%$ of the total distribution and the unfolded state (S1) has 69.5$\%$ of the total distribution which is in great agreement with other studies on this dataset using Markov state models.\cite{sidky2019high, ghorbani2021variational}. GraphVAMPNet produces an embedding for each timestep of the simulation which is then turned into a membership assignment using a softmax function. This higher dimensional embedding (16 dimension) can be visualized using dimensionality reduction methods such as t-SNE (figure S3A) which shows the separated clusters based on their maximum membership assignment probability. To have a better visualization of the low-dimensional space learned by the model, we also trained a GraphVAMPNet where in the last layer we linearly transformed the learned graph embedding into 2-D and trained the model by maximizing the VAMP-2 score. Other parameters were kept similar to the main SchNet. The 2-D free energy landscape (FEL) for this embedding is shown in figure 2C. This low energy states in the FEL correspond to the states with high cluster assignment probability. This low-dimensional FEL shows the ability of the GraphVAMPNet to produce an interpretable and highly clustered embedding of graphs for simulation of proteins. The learned 2-D embedding of graphs during TrpCage Folding is shown in figure 2D where the states with more than 0.95 cluster assignment probability are colored. The enhancement of the SchNet with attention gives an interpretable model where we can analyze the nodes and edges in the graph that are most important in each coarse-grained cluster. The scaled attention scores for TrpCage are shown in figure 3. The cage residue Trp6 shows a high attention score in most clusters due to being in the center of the protein and having a high number of connections in the graph. In the unfolded state (S1) most residues have high attention score only with their close neighbors on the sequence which is due to high level of dynamic and no defined structure of the unfolded state. On the other hand, other clusters such as S2 show different hot spot regions for their attention scores. In this hairpin-like structure, residues Ala4, Ser14 and Pro17 which make the groove have high attention scores. A two step folding mechanism has been proposed for TrpCage that involves an intermediate state with a salt bridge between Asp9 and Arg16.\cite{zhou2003trp} Breaking this salt-bridge is thought to be a limiting step in folding of TrpCage. Surprisingly, our model puts high attention scores on residues Arg16 and Asp9 in metastable state S3 which also has a 10$\%$ probability. 
 
\subsection{Villin}
Villin is a 35-residue protein and is known as one of the smallest proteins that can fold autonomously. It is composed of 3 $\alpha$-helices denoted as helix-1 (residues 4-8), helix-2 (residues 15-18), helix-3 (residues 23-32) and a compact hydrophobic core. The double mutant of Villin with replacement of two Lys residues with uncharged Norleucine (Nle) was simulated by DE Shaw \cite{lindorff2011fast} group and is studied here. Hernandez and coworkers \cite{hernandez2018variational} used a variational dynamic encoder to produce a low-dimensional embedding of Villin folding trajectories using $C_{\alpha}$ distance maps. The optimized TICA for this protein used a lagime of 44 ns according to hyperparameter optimization. \cite{husic2016optimized}
\begin{figure*}
    \centering
    \includegraphics[scale=0.5]{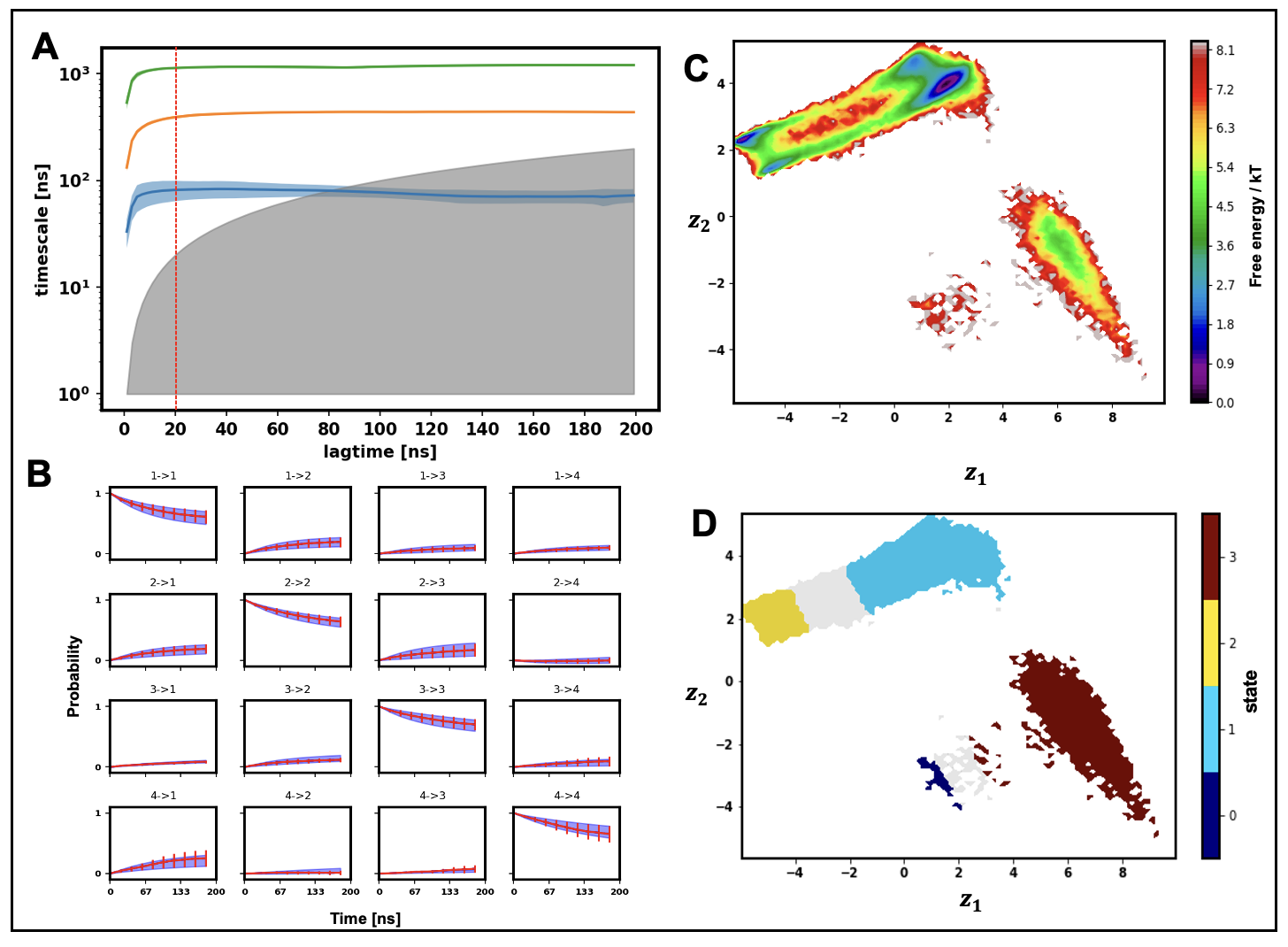}
    \caption{ Villin system A) Implied timescale (ITS) plot for SchNet as feature transformation in VAMPNet (errors are taken from 95 \% confidence interval from 10 different trainings) B) CK-test for SchNet using lagtime of 20 ns C) Free energy landscape (FEL) in a 2d graph embedding F) state assignment of the 2d graph embedding using 0.75 cutoff. }
    \label{fig:my_label}
\end{figure*}

We built VAMPNet with different types of feature functions and compared the average VAMP-2 score of validation set for 10 different training between them. VAMPNet based on SchNet showed the highest VAMP-2 score for the same number of states (4). The VAMP-2 score for training and validation set are shown in figure S1B for SchNet which shows a converging behavior for 100 epochs of training. The VAMPNet built using SchNet shows an extremely fast convergence of implied timescales even after 20 ns (figure 4A) which gives a high resolution kinetic model for villin folding.  On the other hand, VAMPNet built with standard VAMPNet shows a slow convergence for implied timescales where the first timescale converges after 40 ns of lagtime (figure S2). The timescales of the processes are also higher for SchNet than in standard VAMPNet model which also demonstrates the higher accuracy of GraphVAMPNet than VAMPNet with simple distance matrix. The CK test for SchNet (figure 4B) shows excellent markovian behavior of the model built using a lagtime of 20 ns at large timescales $k\tau = 200$ ns. GraphVAMPNet also provides a latent embedding for graphs which is another advantage of GNN features compared with standard VAMPNet layer. The 16-dim graph embedding for Villin was further reduced to 2 dimensions using t-SNE (figure S3B). This 2-D latent embedding shows highly separated clusters. To have a more interpretable embedding we have trained a VAMPNet using SchNet where we linearly transformed the last embedding layer into 2 dimensions and trained the model using similar parameters as before. The 2-D embedding of Villin learned using GraphVAMPNet is shown in figure 4D where datapoints with a cluster assignment probability higher than 0.75 are colored based on their corresponding state. The FEL on this 2-D embedding is shown in figure 4C. This FEL features highly separated clusters with low energy minima corresponding to the center of clusters and the transition regions having low membership assignment probabilities. The representative structure of each cluster of Villin (Misfolded: S0, Unfolded: S1,  Partially-folded: S2, Folded: S3) are shown in figure 5 which are colored based on the average attention score for each residue in that cluster. The folded state (S3) shows a high attention score for residue Arg13. This is in agreement with previous study by Mardt et al.\cite{mardt2021progress} who used distance map features and attention on neighboring residues. We found residues Gln25 and Nle29 to also have high attention scores in the folded state. These residues are in the central hydrophobic core of the protein and have high number of connections in the graphs built for the folded state. The partially folded (S2) state has similar attention scores to the folded state except for residue Lys7 which shows high attention score in partiall folded (S2) but not in folded (S3). The misfolded state (S0) has high attention score for helix 2 residues which is also in agreement with work done by  Mardt et. al.\cite{mardt2021progress} In general the N and C-terminal of the protein due to their high flexibility are given low attention scores. Hernandez and coworkers \cite{hernandez2018variational} used variational dynamic encoders to reduce the complex nonlinear folding of Villin into a single embedding and used a saliency map to find important Ca contacts for folding of Villin. They found residues Lys29 and His27 to be important for the folding of Villin. We found these residues to have high attention scores in our model for partially folded and folded states. 

\begin{figure*}
    \centering
    \includegraphics[scale=0.55]{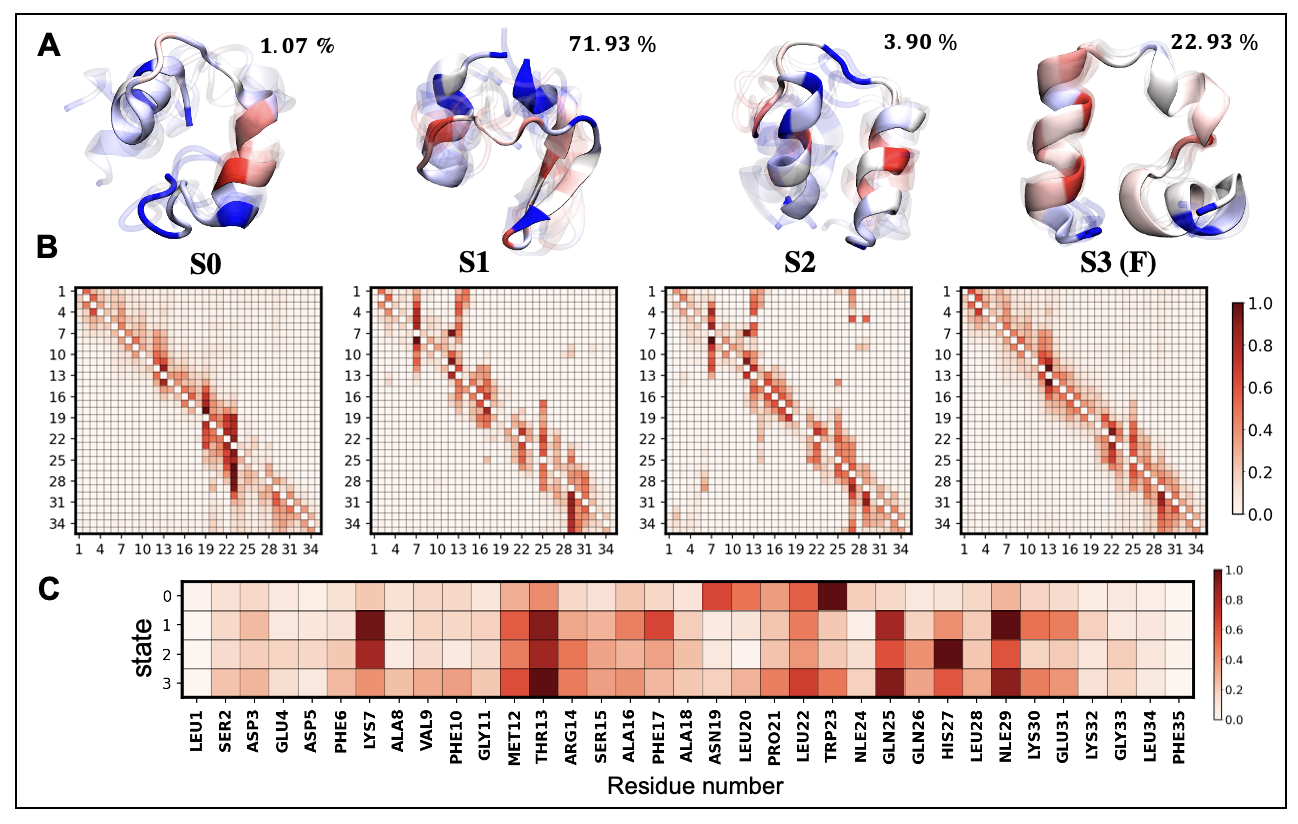}
    \caption{Representative structure of each metastable state in Villin with their probabilities B) average attention score between $C_\alpha$ atoms for each cluster C) averaged attention score for each residue of Villin in each cluster which is the scaled sum of rows.}
    \label{fig:my_label}
\end{figure*}

\subsection{NTL9}

As our last example, we tested the GraphVAMPNet on the NTL9 (residues 1-39) folding dataset from DE Shaw group.\cite{lindorff2011fast}. We uniformly sampled the 1.11 ms trajectory using a lagtime of 5 ns. Mardt et al.\cite{mardt2018vampnets} previously used a 5-layer VAMPNet with contact maps between neighboring heavy-atoms to coarse-grain the NTL9 simulation into metastable states. They showed that the relaxation timescales by a 5-state VAMPNet correspond to a 40-state MSM. Their implied timescales showed a converging behavior after about 320 ns which they chose as the lagtime of the Koopman matrix. 
\begin{figure*}
    \centering
    \includegraphics[scale=0.6]{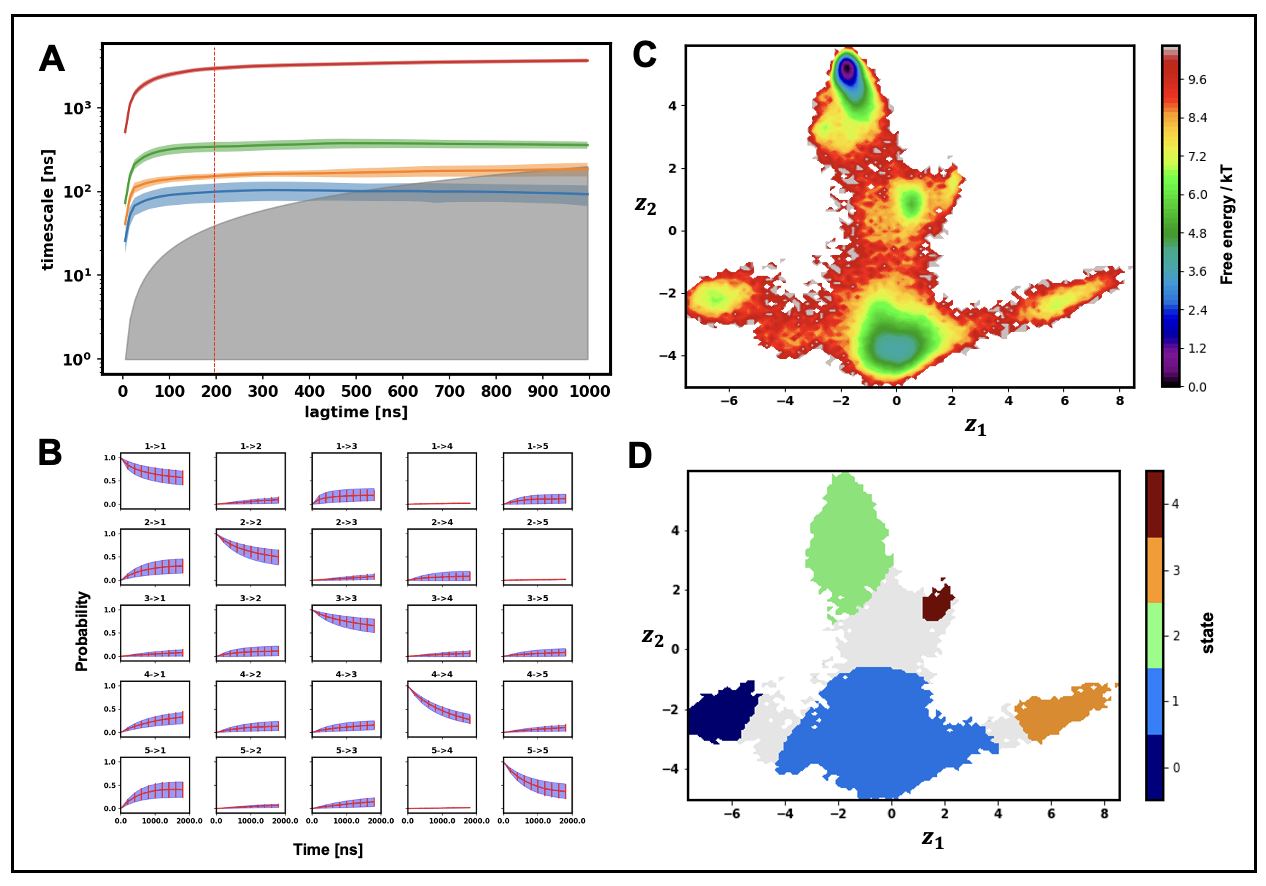}
    \caption{ NTL9 system A)Implied timescale (ITS) plot for SchNet as feature transformation in VAMPNet (errors are taken from 95 \% confidence interval from 10 different trainings) B) CK-test for SchNet using lagtime of 200 ns C) Free energy landscape (FEL) in a 2d graph embedding F) state assignment of the 2d graph embedding using 0.95 cutoff.}
    \label{fig:my_label}
\end{figure*}

The comparison of VAMP-2 score for VAMPNet built using different neural network feature transformations is shown in table S1. Standard VAMPNet based on the distance maps shows a similar VAMP-2 score compared to SchNet based VAMPNet. The training and validation VAMP-2 score for NTL9 is shown in figure S1C which shows a highly converging behavior after 100 epochs. The implied timescales for SchNet layer if shown in figure 6A. The convergence of implied timescale for SchNet (figure 6A) and standard VAMPNet (figure S2F) and the magnitude of the implied timescales are similar. A lagtime of 200 ns was chosen to build the koopman matrix. The CK test (figure 6C) for SchNet using a lagtime of 200 ns shows the markovianity of the model even at high timescales of 2000 ns.  The t-SNE for the 16 dimensional embedding of NTL9 is shown in figure S3C which shows separated clusters. As described for other proteins, we have trained a SchNet based VAMPNet for NTL9 with a 2-D embedding. Figure 6D shows the 2-D embedding of NTL9 which is colored by states higher than 0.95 cluster membership probability. The FEL in this 2-D embedding (figure 6C) shows low energy metastable states separated by transition regions that correspond to points where model is uncertain about their membership. Representative structures of each cluster in NTL9 are shown in figure 7 (colored based on residue attention scores). The folded state (S2) and unfolded state (S0) posses 75.5 and 18.7 $\%$ of the total probability distribution. Schwantes et al.\cite{schwantes2013improvements} used a TICA MSM for NTL9 and showed that the slowest timescale (~18$\mu$s) corresponds to the folding process while the faster timescales correspond to transition between different register-shifted states. Register shift in each strand can also shift the hydrophobic core contacts. For instance based on their study, register shift in strand 3 produces a shift in the core packing in which Phe29 is packed. Interestingly in our model, high attention scores are given to the beta-strand residues such as Phe5, Phe29, Phe31 and Ile18.

\begin{figure*}
    \centering
    \includegraphics[scale=0.45]{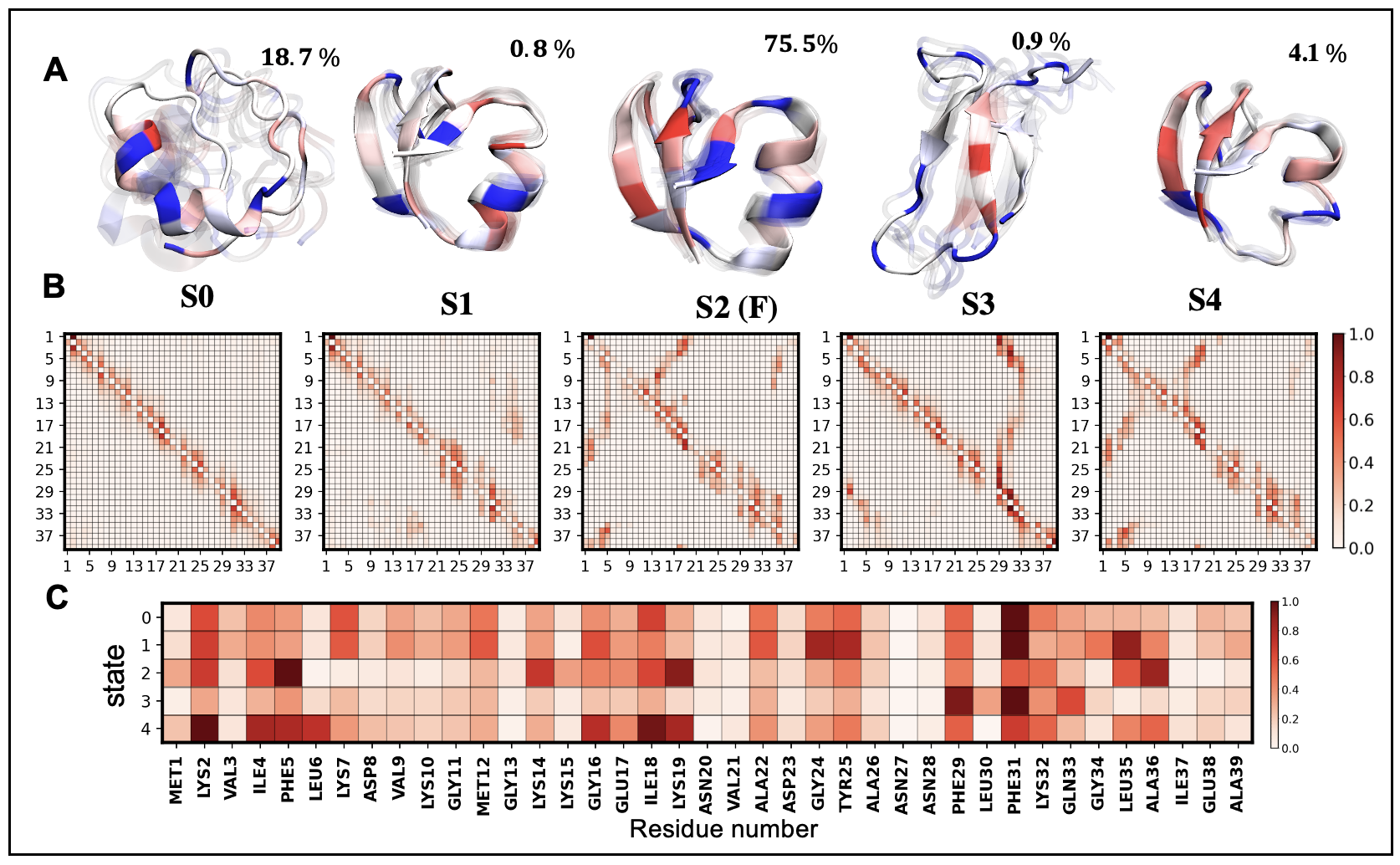}
    \caption{Representative structure of each metastable state in NTL9 with their probabilities B) average attention score between $C_\alpha$ atoms for each cluster C) averaged attention score for each residue of NTL9 in each cluster which is the scaled sum of rows.}
    \label{fig:my_label}
\end{figure*}

\section{Discussion and Conclusion}
MSM construction has previously been a complex process which involved multiple steps such as feature selection, dimension reduction, clustering, estimating the transition matrix K. Each of these steps requires choosing some hyperparameters  and suboptimal choices could lead to poor kinetic model of the system with lower kinetic resolution.\cite{scherer2019variational} Variational approach for conformational dynamics (VAC) and its more general form the variational approach for Markov processes (VAMP) have recently guided the optimal choice of hyperparameters.\cite{wu2017variational, nuske2014variational, scherer2019variational} A cross-validated variational scores is usually used to find the set of features with the highest cross-validated VAMP-2 score.\cite{scherer2019variational}

The end-to-end deep learning framework VAMPNet was proposed by Mardt et al.\cite{mardt2018vampnets} to replace the MSM construction pipeline by training a neural network that maps the molecular configurations x to a fuzzy state space where each point has a membership probability to each of the metastable states. VAMPNet is trained by maximizing a VAMP score allowing us to find the optimal state space which enables linear propagation of states through a transition matrix. VAMPNets are not restricted to stationary and equilibrium MD and can be used as general case for non-stationary and non-equilibrium processes. The few-state coarse-grained MSM in the case of VAMPNet is learned without the loss of model quality as is the case in standard pipelines such as PCCA.\cite{roblitz2013fuzzy} Due to the end-to-end nature of deep neural networks, VAMPNets require less expertise to build an MSM. The framework of VAMPNet were further developed into state-free reversible VAMPNets (SRVs) not to approximate MSMs but rather to directly learn nonlinear approximations to the slowest dynamics modes of a MD system obeying detailed balance.\cite{chen2019nonlinear} In SRVs, the transfer operator rather than soft metastable state assignment, directly employs the variational approach under detailed balance to approximate the slow modes of equilibrium dynamics. Ferguson and coworkers \cite{sidky2019high} showed that MSMs constructed from nonlinear SRV approximations would permit the use of shorter lagtimes and therefore furnish the models with higher kinetic resolution. 

Despite the success of VAMPNet, the feature selection is still a process that must be done with caution. Traditionally distance maps are used as a general feature representation of protein dataset. However this representation does not preserve the graph-like structure of proteins as it does not capture the 3d structure and models the protein as points on a regular grid. In this work we have focused on representation learning of VAMPNet using graph-representation of protein to get a higher-resolution kinetic model where a smaller lag-time can be chosen. Graph representation of molecules is shown to be effective in extracting different properties using deep learning. Recently there has been a large amount of work in the area of geometric deep learning \cite{bronstein2017geometric} that has graph based approaches for representing graph structures. These methods enable automatic learning of the best representation (embedding) from raw data of atoms and bonds for different types of predictions. \cite{kearnes2016molecular, marcheggiani2017encoding} These methods have been applied to various tasks such as molecular feature extraction  \cite{duvenaud2015convolutional, kearnes2016molecular} protein function prediction \cite{gligorijevic2021structure} and protein design \cite{strokach2020fast} to name a few. Park et al.\cite{park2021accurate}  proposed a machine learning framework (GNNFF) a graph neural network  to predict atomic forces from local environments and showed its high performance in force prediction accuracy and speed.   Introduction of graph message passing enhances the model ability to recognize symmetries and invariances (permutation, rotation and translation) in the system. Hierarchical pooling from atom-level to residue level and then to the protein level enables the model to learn global transition between different metastable states that involves atomic-scale detailed dynamics. Xie and coworkers \cite{xie2019graph} developed graph dynamical networks (GdyNet) to investigate atomic scale dynamics in material systems where each atom or node in the graph has a membership probability to the metastable states. Graph representation of materials in their model enabled encoding of local environment that is permutation, rotation and reflection invariant. The symmetry in materials facilitated identifying similar local environment throughout the material and learning of the local dynamics. This type of approach can be used to learn local dynamics in some biophysical problems such as nucleation and aggregation where local environment is important. 

The introduction of Graph Neural networks to VAMPNets enables the higher resolution of the kinetic model and higher interpretability. A large increase in VAMP-2 score is observed when switching from distance based features to graph-based. This suggests the usefullness and representation capability of GNN for further improving the kinetic embedding of MD simulation. We have tested the GraphVAMPNet with two different types of graph neural networks (Graph Convolution layer and SchNet) on three long-timescale protein folding trajectories. The GraphVAMPNet showed a higher VAMP-2 score than the standard VAMPNet and the implied timescales showed a faster convergence in GraphVAMPNet due to an efficient representation learning as opposed to standard VAMPNet. This enables choosing a lower lagtime for building the dynamical model and improves the kinetic resolution of the resulting Markov model. The graph embeddings resulted from GraphVAMPNet are highly interpretable and shows clustered data in low energy minima in a free energy landscape. Furthermore, the addition of attention mechanism into SchNet enables us to decipher the residues and bonds that most contribute to each of the metastable states. However care must be taken when interpreting the attention scores returned by the model. One main obstacle of GNNs is that they cannot go deeper than a few layers (3 or 4) and suffer from over-smoothing problems in which all node representations tend to become similar to one another. An architecture that enables training deeper networks is the residual or skip connections as deployed in ResNet architecture is used here to train a deep neural network.\cite{he2016deep}

Due to the flexibility of graph representation of molecules, other physical properties of atoms or amino acids such as electric charge, hydrophobicity. can be encoded into node or edge features in order to enhance the physical and chemical interpretability of the model. Moreover, hierarchical pooling layers can be applied to learn the dynamics at different resolutions of the molecule.\cite{ying2018hierarchical} Our GraphVAMPNet is inherently transferable. This means theoretically given a sufficient amount of dynamical data, transfer learning can be leveraged to reduce the number of trajectories needed for studying the dynamics of a particular system of interest. 

Time-reversibility and stochasticity of the transition matrix are the two physical constraints that are needed to get a valid symmetric transition matrix for analyses such as transition path theory (TPT).  Physical constraints added to the VAMPNet and the resulting model called revDMSM was successfully applied to study kinetics of disordered proteins. This allowed to have a valid transition matrix and therefore rates could be quantified for interesting processes. This revDMSM was further extended by including experimental observables into the model as well as a novel hierarchical coarse-graining to give different levels of detail. These physical constraints can be further added into GraphVAMPNet to obtain a valid and high resolution transition matrix.

In summary, our GraphVAMPNet automates the feature selection in VAMPNet to be learned from graph message passing on the molecular graphs which is a general approach to understand the coarse-grained dynamics. 

\subsection*{Acknowledgements}
This work was partially supported by the National Heart, Lung and Blood institute at the National Institute of Health for B.R.B. and M.G.; in addition, it was partially supported by the National Science Foundation (grant number CHE-2029900) to J.B.K. The authors acknowledge the biowulf high-performance computation center at National Institutes of Health and Lobos for providing the time and resources for this project. We also would like to thank DE. Shaw research group for providing the simulation trajectories.

\subsection*{Conflicts of Interest}
The authors declare that there is no conflict of interest regarding the publication of this article.

\subsection*{Data Availability}
The source code for the analysis can be found at github page:
github.com/ghorbanimahdi73/GraphVampNet

\section*{Supplementary Materials}
The training and validation loss and the results of GraphVAMPNet model with  for Trp-cage,  Villin and NTL9 can be found in the supplementary information.

\section*{References}
\bibliography{mybib.bib}
\end{document}



\title{ Supplementary Information for GraphVAMPNet, using graph neural networks and variational approach to markov processes for dynamical modeling of biomolecules}

\author{Mahdi Ghorbani}
\email{ghorbani.mahdi73@gmail.com.}
\affiliation{ 
Laboratory of Computational Biology, National, Heart, Lung and Blood Institute, National Institutes of Health, Bethesda, Maryland 20824, USA.
}%
 \affiliation{Department of Chemical and Biomolecular Engineering, University of Maryland, College Park, MD 20742, USA}
 
\author{Samarjeet Prasad}%
\affiliation{ 
Laboratory of Computational Biology, National, Heart, Lung and Blood Institute, National Institutes of Health, Bethesda, Maryland 20824, USA.
}%

\author{Jeffery B. Klauda}
\affiliation{%
Department of Chemical and Biomolecular Engineering, University of Maryland, College Park, MD 20742, USA}%
\author{Bernard R. Brooks}%
\affiliation{ 
Laboratory of Computational Biology, National, Heart, Lung and Blood Institute, National Institutes of Health, Bethesda, Maryland 20824, USA.
}%

\date{\today}

\maketitle

\maketitle

\begin{table*}
\caption{Average VAMP-2 score for each system}
\begin{tabular}{|c|c|c|c|}
\hline
\textbf{system}  & \textbf{Standard VAMPNet} & \textbf{GraphConvLayer} & \textbf{SchNet} \\ \hline
\textbf{TrpCage} & 4.68 $\pm$ 0.08           & 4.76 $\pm$ 0.03         & 4.79 $\pm$ 0.01 \\ \hline
\textbf{Villin}  & 3.74 $\pm$ 0.02           & 3.74 $\pm$ 0.06         & 3.78 $\pm$ 0.02 \\ \hline
\textbf{NTL9}    & 4.67 $\pm$ 0.03           & 4.27 $\pm$ 0.63         & 4.59 $\pm$ 0.09 \\ \hline
\end{tabular}
\end{table*}

\begin{figure}
    \centering
    \includegraphics[scale=0.65]{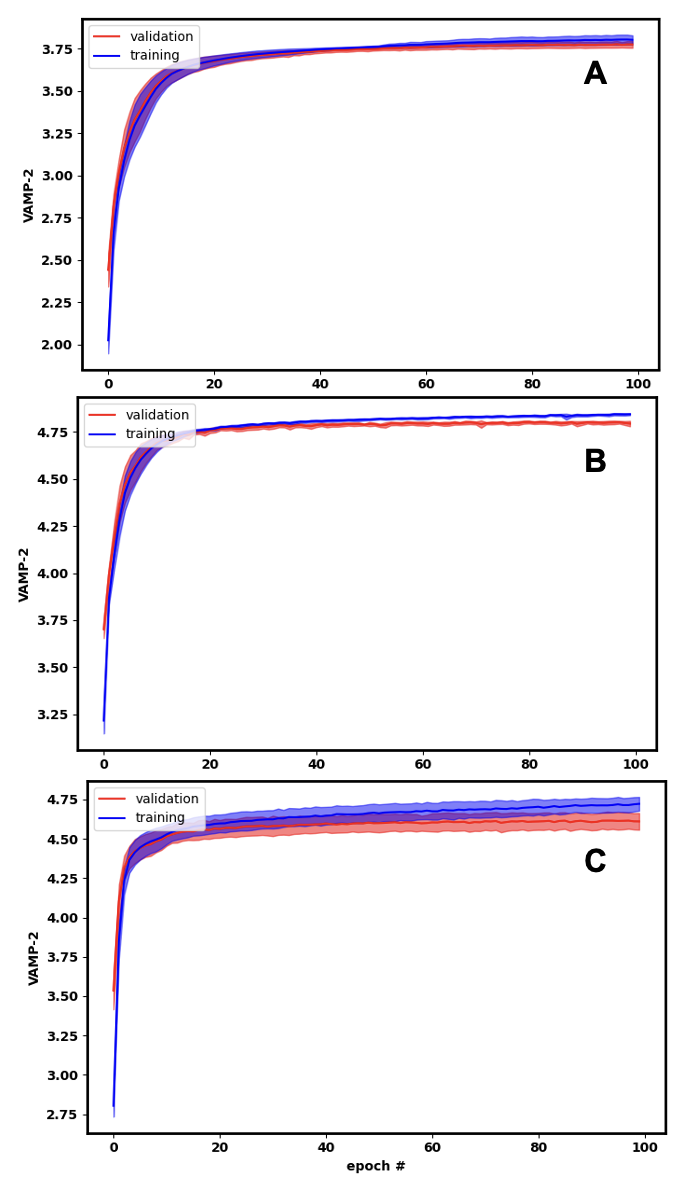}
    \caption{training and validation losses from SchNet based VAMPNet for A)Trp-cage B)Villin C)NTL9}
    \label{fig:my_label}
\end{figure}

\begin{figure}
    \centering
    \includegraphics[scale=0.65]{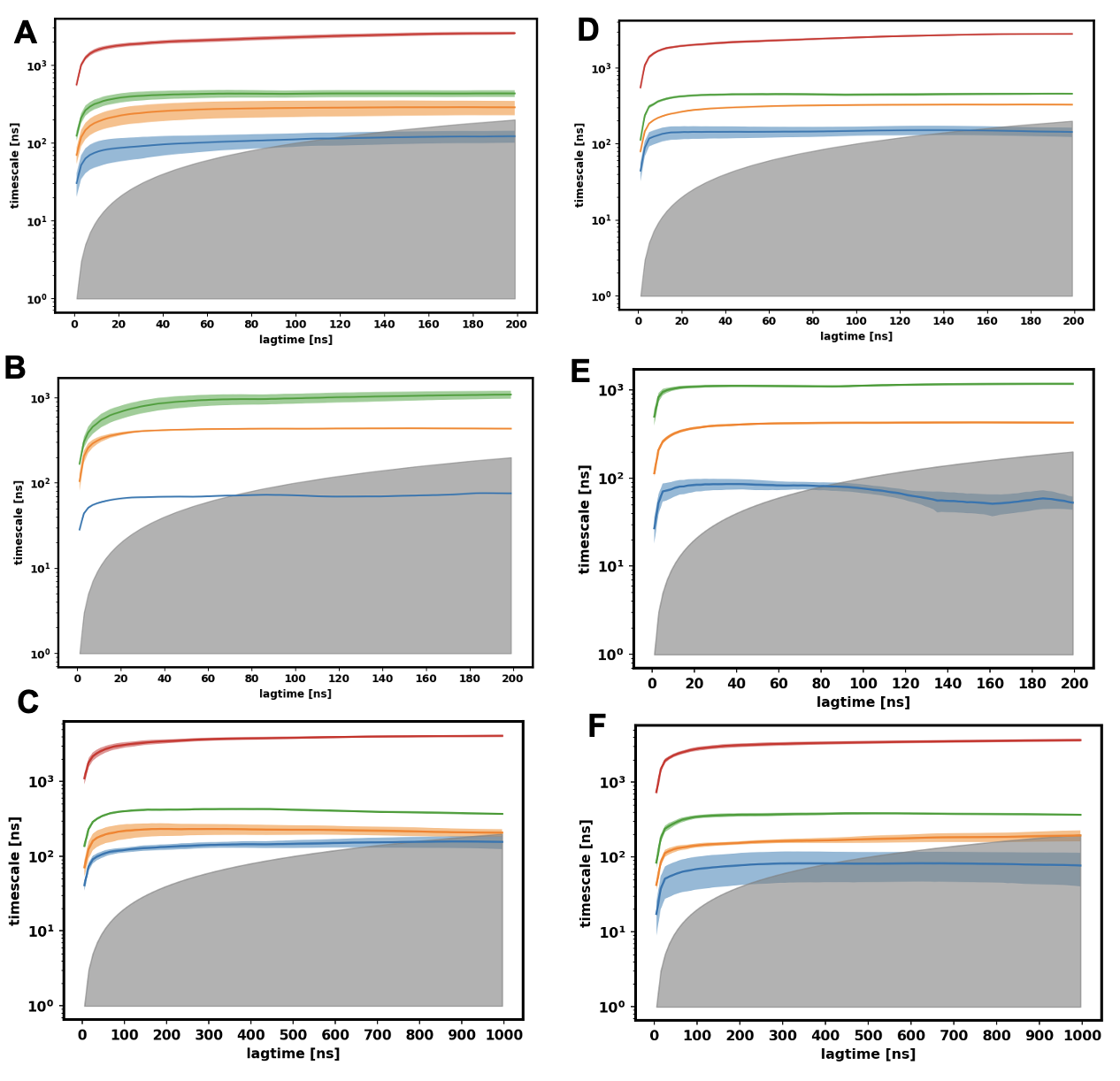}
    \caption{Implied timescales for standard VAMPNet A)TrpCage B)Villin C)NTL9 and VAMPNet based on Graph Convolution layer for D)TrpCage E)Vilin F)NTL9}
    \label{fig:my_label}
\end{figure}

\begin{figure}
    \centering
    \includegraphics[scale=0.65]{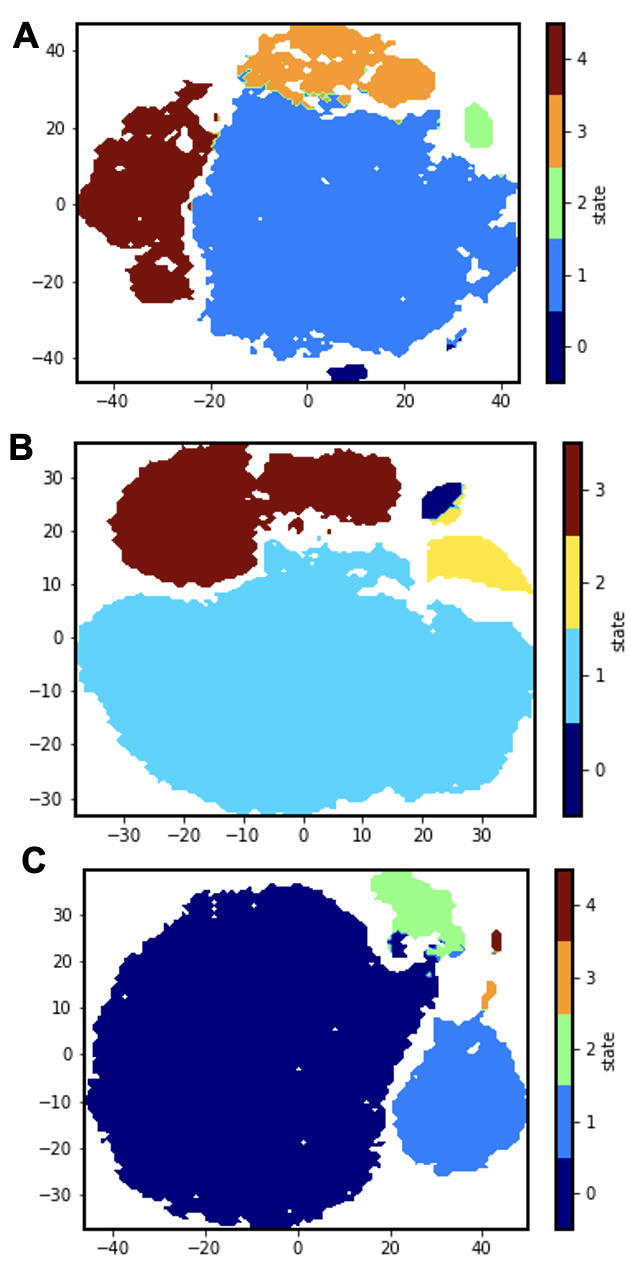}
    \caption{TSNE visulation of 16d graph embedding of SchNet based VAMPNet for A)TrpCage B)Villin and C)NTL9}
    \label{fig:my_label}
\end{figure}